\documentclass[a4paper]{panl}
\usepackage{cite}
\usepackage{wrapfig}
\usepackage{graphicx,multicol}
\usepackage{amssymb}
\usepackage{amsfonts}
\usepackage{amsmath}
\usepackage{longtable}
\usepackage{rotating}
\usepackage{lscape}
\usepackage{epsfig}
\usepackage{lineno,hyperref}
\usepackage{subfig}
\usepackage{float}
\usepackage{array}
\usepackage{csquotes}
\usepackage{titlesec}

\originalTeX
\begin{document}

\title{To the UCN source with pulsed filling of a trap}
\maketitle
\authors{A.I.\,Frank$^{a,b,}$\footnote{E-mail: frank@jinr.ru},
G.V.\,Kulin$^{a,b,}$\footnote{E-mail: kulin@jinr.ru}, M.A.\,Zakharov$^{a,b}$, S.V.\,Mironov$^{a}$}
\authors{V.A.\,Kurylev$^{a,b}$, A.A.\,Popov$^{a,b}$, K.S.\,Osipenko$^{a,b}$}
\setcounter{footnote}{0}

\from{$^{a}$\, Joint Institute for Nuclear Research, Joliot-Curie str. 6, 141980, Dubna, Russia}
\from{$^{b}$\, Dubna State University, Universitetskaya str. 19, 141980, Dubna, Russia}

\begin{abstract}
The paper is devoted to the discussion of the possibility of creating UCN sources based on the principle of pulse accumulation (PA) in traps. The implementation of the PA principle would make it possible to create a source with a flux of UCN in a trap significantly exceeding the time average. The paper provides a comparative analysis of various approaches to the implementation of the idea of PA of UCN in traps remoted from the place of their generation. Based on this analysis, the concept of the UCN source, the creation of which is planned at the IBR-2M pulse reactor, was formulated. A distinctive feature of the designed source is a combination of several approaches to ensuring the pulsed structure of neutron bunches reaching the UCN trap. One of them is the deceleration of the pulsed flux of VCN using a resonant flipper, the second is the use of compensating time lenses.
\end{abstract}
\vspace*{6pt}


\section{\uppercase{Introduction}}
\label{sec:intro}

Ultracold neutrons were first observed by Shapiro's group in an experiment performed on a pulsed reactor with an average power of 6 kW \cite{Luschikov69, Strelkov15}. The subsequent development of UCN physics was associated with the creation and use of UCN sources at stationary reactors \cite{Groshev71, Egorov74, AltarevPLA80, Altarev86, Steyerl86}. Later on, a UCN source on a pulsed reactor with rare pulses \cite{Karch14}, as well as sources based on proton accelerators \cite{Anghel09, Saunders13}, appeared. In such reactors, a UCN trap could be filled over a relatively short period of time, but the interval between the filling cycles was comparable to the lifetime of neutrons in the trap.

At the same time, there are pulsed neutron sources being built and designed, which have a relatively high repetition rate and a short (milliseconds or less) pulse. They include the IBR-2M reactor, which has been in operation for a long time \cite{Ananiev77, Aksenov09}, the European source ESS \cite{Garoby18}, China Spallation Neutron Source \cite{Chen16} and the Neptune reactor being designed \cite{Lopatkin21}. Since the pulse repetition period of such sources exceeds their pulse duration considerably, they have a very high pulse density of the neutron flux. It significantly exceeds not only the time-averaged value, but also the flux achievable in modern high-flux reactors. Therefore, the question of how to take advantage of this circumstance when creating a UCN source is rather relevant, although not new.

A consistent approach to solving this problem was described in the works of A. V. Antonov and his colleagues \cite{Antonov69} and, in more detail, by F.L. Shapiro \cite{Shapiro71, Shapiro74}. It consists in filling the trap with UCNs only during the pulse and effectively isolating the trap for the rest of the time. If during the source operation period $T$ the number of neutrons leaving the trap for various reasons is less than the number of neutrons entering the trap during the pulse, the neutron density in the trap will increase until it reaches some equilibrium value. Ideally, when there are no losses, the UCN density in the trap may correspond to a pulsed neutron density significantly exceeding the average one in time.

The practical implementation of this idea is hindered by the fact that, due to the presence of biological shielding, the trap is remote from the moderator, where UCNs are born. In this case, there is a need for a transport neutron guide several meters long to feed the trap. The placement of an insulating valve near the moderator – the source of UCNs – leads to the neutron guide becoming part of this trap. Due to a small transverse size of the neutron guide, the frequency of neutron collisions against its walls is quite high, which greatly reduces the time of UCN storage in the trap-neutron guide system and significantly reduces the density of neutrons accumulated in the trap. Placing the valve at the entrance to the trap also makes no sense, since the dispersion of the UCN times of flight from the source to the trap (for sources with a repetition rate of several Hertz) will substantially exceed the intervals between the pulses.

Therefore, to create a UCN source with pulsed accumulation in a trap, it is necessary, in any way, to ensure conditions for a short duration of neutron bunches entering a trap that is remote from the place where neutrons are born. This paper is devoted to discussion of possible ways to solve this problem.

The structure of the paper is as follows. The next section provides a simple theory of pulse accumulation of UCN in a trap. The third section of the article is devoted to a well-known method for restoring the initial pulse structure of the UCN flux at a point far from the neutron generation site. It is based on a time–dependent change in the kinetic energy of neutrons by a special device - a time lens. An elementary theory of time focusing is presented, and the most important question of the ratio of the initial and final duration of a time pulse is discussed. The fourth section is devoted to a comparative analysis of known time focusing methods.

Relatively recently, a new possibility of pulsed accumulation of UCN has been found \cite{FrankPEPANL23}, based on the deceleration of very cold neutrons (VCN) by a local decelerating device. The fifth section of the article is devoted to its presentation. Two fundamentally different approaches to neutron deceleration are discussed in two parts of this section. The sixth section of the article is devoted to the possibility of combining time focusing and the method of local deceleration. The review part of the article concludes with a short section on the problem of a pulse valve, which is an obligatory component of a UCN source with a pulse trap accumulation.

The eighth section contains a brief comparative analysis of the above-discussed possibilities for creating a pulse storage UCN source and describes the concept of an UCN source on an existing IBR-2M reactor. The main results of the work are summarized in the Conclusion.

\section{\uppercase{Pulse accumulation of UCN in a trap}}
\label{sec:sec2}

Let us now clarify the idea of the accumulation of neutrons from a pulsed source, following the work by F. L. Shapiro  \cite{Shapiro74}. Let us have a trap of ultracold neutrons, and a periodic pulsed flux with pulse duration $\tau$ enters the entrance hole of area $s$. Then, the number of neutrons entering the trap in a pulse is

\begin{linenomath}\begin{equation}\label{eq:eq8}
N_{in}=F_s s \tau,
\end{equation}\end{linenomath}
where $F_s$ is the neutron flux averaged over the pulse duration. Suppose that the entrance window of the trap is open only during the pulse, and the rest of the time is blocked by some ideal shutter. The equilibrium value of the flux in the trap $\Phi$ is easy to calculate. Neglecting the probability of neutron decay during the time of the storage in the trap, we assume that UCNs can escape from the trap only by being absorbed in its walls or through the entrance hole during the time when it is open. Equating the number of neutrons entering the trap in one pulse~\eqref{eq:eq8} to the number of neutrons leaving it during the duration of one cycle $T$, we obtain

\begin{linenomath}\begin{equation}\label{eq:eq9}
F_s s\tau=\Phi\left(s\tau+\Sigma T\mu \right),
\end{equation}\end{linenomath}
where $\Sigma$ is the area of the trap surface, $\mu$ is the UCN absorption coefficient in a collision with the trap wall. For the equilibrium flux, we have
\begin{linenomath}\begin{equation}\label{eq:eq10}
\Phi=F_s\frac{1}{1+\left(\Sigma\mu T/s\tau\right)}.
\end{equation}\end{linenomath}

From formula~\eqref{eq:eq10}, it is easy to obtain expressions for the ratio of the flux in the trap $\Phi$ to the mean input flux of the trap $\langle F\rangle=F_s\tau/T$, that is, the gain factor.

\begin{linenomath}\begin{equation}\label{eq:eq11}
G=\frac{\Phi}{\langle F\rangle},\;\;\;G=\frac{sT}{s\tau+\Sigma\mu T}.
\end{equation}\end{linenomath}

Formula~\eqref{eq:eq11} slightly differs from formula (8) in \cite{Shapiro74}, which is associated with two circumstances. First, the possibility of neutron outflow from the trap to the user is not considered in~\eqref{eq:eq11}. Secondly, in \cite{Shapiro74}, the neutron flux in the trap is compared in the case of pulsed and stationary filling methods, rather than the ratio of flux in the source and trap.

For a trap of a significant volume with weakly absorbing walls, the storage time may be so long that neglecting neutron decay turns out to be wrong. To take this into account, one should add the expression $N_\gamma=nQ\left(1-e^{-\gamma T}\right)$ to the right side of equation~\eqref{eq:eq9}, where $n=2\Phi/\langle v\rangle$ is the density of neutrons in the trap, $\langle v\rangle$ is their average velocity, $Q$ is the volume of the trap, and $\gamma$ is the decay constant. Since $\gamma T\ll1$, one can assume that $N_\gamma=nQ\gamma T$.   Hence, instead of~\eqref{eq:eq10} we get

\begin{linenomath}\begin{equation}\label{eq:eq12}
\Phi=F_s\frac{s\tau}{\left(s\tau+\Sigma T\mu+2Q\gamma T/\langle v\rangle\right)}.
\end{equation}\end{linenomath}

The absorption coefficient of neutrons reflected from the wall $\mu$ is determined by the ratio of the imaginary and real parts of the effective potential $U$ of interaction between neutrons and the trap material: $\eta=W/E_b$ \cite{Shapiro74, Ignatovich90}.
\begin{linenomath}\begin{equation}\label{eq:eq13}
U=E_b-iW,\;\;\;E_b=\frac{2\pi \hbar^2}{m}\rho b,\;\;\;W=\frac{\hbar}{2}\rho\sigma v.
\end{equation}\end{linenomath}

Here, $b$ is the neutron scattering length on the nuclei of the trap material, $\rho$ is the number of nuclei per volume unit, $\sigma$ is the cross section of all processes in the trap material that lead to the loss of UCNs, and v is the neutron velocity. The real part of the potential $E_b$ is usually called the boundary energy of substance.

For an isotropic flux, the angular-averaged absorption coefficient $\mu$ is
\begin{linenomath}\begin{equation}\label{eq:eq14}
\mu=\frac{2\eta}{y^2}\left(\arcsin y-y\sqrt{1-y^2}\right),
\end{equation}\end{linenomath}
where $y=\sqrt{E/E_b}$, and $E$ is neutron energy \cite{Shapiro74, Ignatovich90}. If the energy $E$ is not too close to the boundary energy, the values $\mu$ and $\eta$ in a wide range of variation of the $y$ parameter differ inconsiderably. For a number of good materials, the value $\eta$ can be $\eta=(3\div5)\times10^{-5}$  \cite{Nesvizhevskii92, Arzumanov03, Atchison07} and the ratio $\Sigma\mu T/s\tau$ of the values in the denominator of Eq.~\eqref{eq:eq10} can be significantly less than unity, even for a trap of quite a significant volume. At the same time, neglecting the neutron decay, the gain factor $G$ turns out to be of the order of $T/\tau$.

It follows from the above-mentioned that the duration of the neutron pulse $\tau$ at the entrance to the trap is one of the most important factors determining the effectiveness of the entire device. 

\section{\uppercase{Neutron focusing in time to restore the pulse structure of the beam}}
\label{sec:sec3}

\subsection{\textit{Time lens operation principle}}
\label{sec:ssec3p1}

Apparently, the question of the possibility to restore the pulsed structure of a neutron beam in a region remote from the moderator was first discussed in \cite{Frank00}. It was proposed to use a special device – a time lens that doses changes in the energy of neutrons as they enter the lens.

Before we turn to the presentation of the time lens operation principle, two important remarks should be made. In the literature, the term "time lens" is used in relation to devices with significantly different functions and operating principles. If in \cite{Frank00} it was a lens forming a given time distribution of the neutron flux at a known point in space, in \cite{Rauch85, Summhammer86, Baumann05} the possibility of compressing the velocity interval at the observation point, that is, additional monochromatization of the neutron beam, was discussed. In addition, \cite{Frank00} considered devices that affect the energy of neutrons, being localized in a relatively small area compared to the total flight length. On the contrary, in \cite{Rauch85, Summhammer86, Baumann05} it was proposed to control the motion of a neutron by interacting with a potential extended in space. Further on, we will mean devices of the first type, which we will call local time lenses for short.

\begin{figure}[htb]
\centering
        \includegraphics[width=0.8\linewidth]{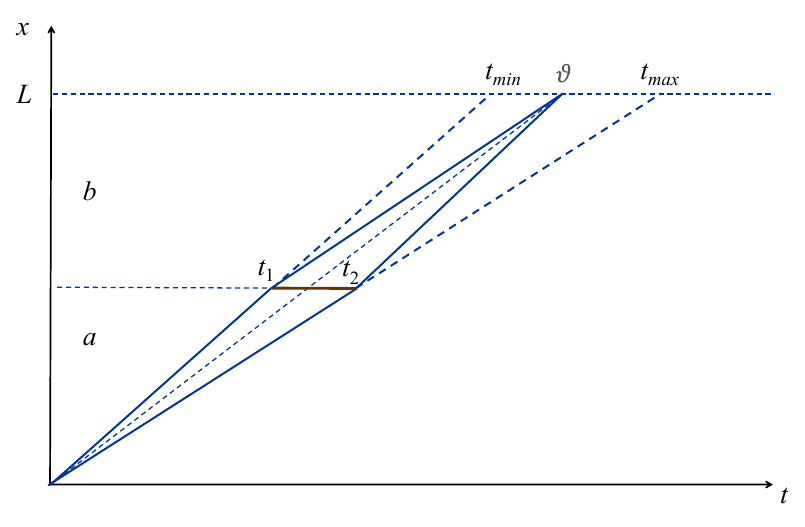}
\caption{Scheme of the time lens operation principle}
\label{fig:Fig1}
\end{figure}

Following the work  \cite{Frank00} (see Fig. 1), let us explain the operation principle of the local time lens. Let neutrons be emitted from a point $x=0$ in the positive direction of the $X$-axis at the time instant $t=0$; their velocities $v$ are distributed in a certain range of values. The time of their arrival $t_L$ at the observation point $x=L$ is distributed in the interval $t_{min}<t_L<t_{max}$. Suppose that a time lens capable of changing the neutron energy by a value $\Delta E$ according to a given time law in the time interval $t_1<t<t_2$ is located at a point $x = a$. Then, the dependence $\Delta E(t)$ can be chosen in such a way that the neutron velocities after neutrons passed through the lens satisfy the condition of simultaneous arrival at the observation point at the instant of time $t_L=\vartheta$.

\begin{linenomath}\begin{equation}\label{eq:eq1}
\frac{a}{V_a}+\frac{b}{V_b}=\vartheta,\;\;\;a+b=L,
\end{equation}\end{linenomath}
where $V_a$ and $V_b$  are neutron velocities before and after their passing through the lens, respectively.          
\begin{linenomath}\begin{equation}\label{eq:eq2}
\Delta E(t)=\frac{m}{2}\left[\left(\frac{b}{\vartheta-t}\right)^2-\left(\frac{a}{t}\right)^2\right],\;\;\;t=\frac{a}{V_a}\;\;\;(t_1<t<t_2),
\end{equation}\end{linenomath}
and $m$ is the neutron mass.

Let us assume that the period of the lens action $T=t_2-t_1$ coincides with the period of source pulses repetition. In this case, the period $T$, the distance $a$ from the source to the lens, and the range of accepted neutron velocities $V_{a\;min}\leq V\leq V_{a\;max}$ are related.

\begin{linenomath}\begin{equation}\label{eq:eq3}
T=\frac{a}{V_{a\;min}}-\frac{a}{V_{a\;max}}.
\end{equation}\end{linenomath}

The distance $a$ and the maximum velocity $V_{a\;max}$ captured by the lens determine the minimum time of flight from the source to the lens

\begin{linenomath}\begin{equation}\label{eq:eq4}
t_{a\;min}=t_1=\frac{a}{V_{a\;max}}.
\end{equation}\end{linenomath}

The minimum velocity captured by the lens and the maximum flight time of section a are determined as follows:
     
\begin{linenomath}\begin{equation}\label{eq:eq5}
V_{a\;min}=\frac{a V_{a\;max}}{TV_{a\;max}+a},\;\;\;t_2=\frac{a}{V_{a\;min}}.
\end{equation}\end{linenomath}

The velocities $V_{a\;max}$ and $V_{a\;min}$ determine the value of the neutron flux transformed by the lens. The range of captured velocities $\Delta V_a=V_{a\;max}-V_{a\;min}$ can be called the velocity aperture of the lens. The value of the maximum neutron velocity after the lens $V_{b\;max}$ is limited by the properties of the trap that accumulates neutrons. By setting this velocity, the minimum flight time of the second section of the neutron guideline $t_{b\;min}=b/V_{b\;max}$  is also determined. The neutrons that spent most of the time in the first section of the flight should have the highest velocities after passing through the lens. Therefore, the total time of flight $\vartheta$ is

\begin{linenomath}\begin{equation}\label{eq:eq6}
\vartheta=t_{a\;max}+t_{b\;min}=\frac{a}{V_{a\;min}}+\frac{b}{V_{b\;max}}.
\end{equation}\end{linenomath}

The maximum energy transfer determined by the velocities $V_{a\;max}$ and $V_{b\;min}$ is

\begin{linenomath}\begin{equation}\label{eq:eq7}
\Delta E_{max}=\frac{m}{2}\left[V_{a\;max}^2-V_{b\;min}^2\right].
\end{equation}\end{linenomath}
Possible ways of changing the neutron energy in accordance with a given time law~\eqref{eq:eq2} will be discussed below in  section \ref{sec:sec4}.

Similar to the way the image formation is associated with the transformation of the angular distribution of rays in optics, time focusing is accompanied by a change in the distribution of neutrons by velocities. At the same time, the duration of the time pulse is transformed. The question of the ratio between the pulse durations of the source $\theta$ and its time image $\tau$ is very important, since the ratio $T/\tau$ defines the relation between the values of the average and pulse fluxes of UCNs at the point $x=L$ where the trap is located. The value $M=\tau/\theta$ will be called a time magnification. The issue of time magnification and duration of the time image is considered in \cite{FrankNIMA23}. Below we will discuss the main points of this work.

\subsection{\textit{Duration of the pulse generated by the time lens}}
\label{sec:ssec3p2}

Let us assume that the source generates neutron flux pulses whose shape is determined by a certain function $f(t)$ with a maximum at $t=0$. Considering the case of a single pulse, we do not pay any attention here to the cyclic nature of the source operation, assuming that the duration of the time image is much less than the time between pulses $T$.

In accordance with the above-stated, a time lens located at a point $x=a$ affects, one way or another, the neutrons reaching it, according to the time law~\eqref{eq:eq2}, which can be conveniently written in the form

 \setcounter{equation}{8}
\begin{subequations}
\begin{linenomath}\begin{equation}\label{eq:eq2a}
\left(V_a^2-V_b^2\right)=\left(\frac{a}{t}\right)^2-\left(\frac{b}{\vartheta-t}\right)^2.
\end{equation}\end{linenomath}
\end{subequations}
 \setcounter{equation}{14}
 
 Condition~\eqref{eq:eq2a} is valid for any neutrons that have reached the lens at time $t$, regardless of the velocity value $V_a$. However, it is obtained under the assumption that neutrons were born at time $t=0$ and the neutron velocity in the first section of the flight path is $V_a=a/t$. For this calculated velocity, as well as for the corresponding velocity on section $b$, we introduce the notation $\tilde{V}_a$ and $\tilde{V}_b$. It is obvious that in the ideal case of zero pulse duration
\begin{linenomath}\begin{equation}\label{eq:eq16}
\tilde{V}_b=\frac{b}{\vartheta-t},\;\;\;\vartheta=\frac{b}{\tilde{V}_b}+\frac{a}{\tilde{V}_a},
\end{equation}\end{linenomath}
which is the condition of focusing. In the case of finite duration $\theta$ of the neutron pulse, neutrons can be born at any time moment, $-\theta/2<t<\theta/2$. Therefore, the velocity $V_a$ in eq.~\eqref{eq:eq2a} may differ from $\tilde{V}_a$. 

Let us calculate the total flight time of a neutron born at an arbitrary moment of time $\delta$. If it reaches the lens at the moment of time $t$, the time of its flight along the path $a$ is $t_a=t-\delta$, and the velocity upon reaching the lens is
\begin{linenomath}\begin{equation}\label{eq:eq17}
V_a=\frac{a}{t-\delta}.
\end{equation}\end{linenomath}

Assuming that in real conditions the magnitude of $a$ is several meters, and the speed of UCNs is about 10 m/s, we get that the time $t$ is about a second. The value of $\delta$ is of the same order as the pulse duration and for periodic pulsed reactors can hardly exceed the value of several hundred microseconds. Therefore, hereinafter we will assume that $\delta/t\ll 1$. Consequently, for speed $V_a$ we have
\begin{linenomath}\begin{equation}\label{eq:eq18}
V_a\approx\tilde{V}_a\left( 1+\frac{\delta}{t}\right),\;\;\;\delta/t\ll 1.
\end{equation}\end{linenomath}

From~\eqref{eq:eq2a} and~\eqref{eq:eq16} it follows that the neutron velocity in section $b$ is
\begin{linenomath}\begin{equation}\label{eq:eq19}
V_b=\sqrt{V_a^2-\left(\frac{a}{t}\right)^2+\left(\frac{b}{\vartheta-t}\right)^2}=\sqrt{V_a^2-\tilde{V}_a^2+\tilde{V}_b^2}.
\end{equation}\end{linenomath}

Substituting~\eqref{eq:eq18} here, we get
\begin{linenomath}\begin{equation}\label{eq:eq20}
V_b=\tilde{V}_b\left(1+\frac{2\delta}{t}\frac{\left(\tilde{V}_a\right)^2}{\left(\tilde{V}_b\right)^2}\right)^{1/2},
\end{equation}\end{linenomath}
and the time of flight of this section is
\begin{linenomath}\begin{equation}\label{eq:eq21}
\vartheta_b=\frac{b}{V_b}=\frac{b}{\tilde{V}_b}\left(1+\frac{2\delta}{t}\frac{\left(\tilde{V}_a\right)^2}{\left(\tilde{V}_b\right)^2}\right)^{-1/2}.
\end{equation}\end{linenomath}

The difference between the time $\vartheta_b$ and the calculated value $\tilde{\vartheta}_b=b/\tilde{V}_b$ is
\begin{linenomath}\begin{equation}\label{eq:eq22}
\xi (t)=\frac{b}{\tilde{V}_b}\left[\left(1+\frac{2\delta}{t}\frac{\left(\tilde{V}_a\right)^2}{\left(\tilde{V}_b\right)^2}\right)^{-1/2}-1\right],
\end{equation}\end{linenomath}
and the value of the term $\left(\tilde{V}_a^2/\tilde{V}_b^2\right)$ depends, generally speaking, on the operation mode of the lens.

It can be seen from eq.~\eqref{eq:eq22} that the arrival moment of the neutron at the observation point $x = b$ depends not only on the moment of its departure $\delta$, but also on the ratio $\varsigma (t)=\tilde{V}_a/\tilde{V}_b$ of the calculated velocities depending, generally speaking, on time
\begin{linenomath}\begin{equation}\label{eq:eq23}
\varsigma (t)=\frac{a}{b}\left(\frac{\vartheta}{t}-1\right)\;\;\;(\vartheta>t).
\end{equation}\end{linenomath}

By choosing the values of $a$, $b$, $t$, and $\vartheta$, it is possible, in accordance with eqs.~\eqref{eq:eq21},~\eqref{eq:eq22}, to completely determine the relationship between the moment of neutron production $\delta$ and the moment of its registration $\vartheta+\xi$. With a given distribution of the source pulse shape $f(t)$, that is, the departure moments $\delta$, it is possible to calculate the shape of the distribution of time delays $\xi$, that is, the shape and duration of the time image pulse. In the general case, one should also take into account the spectral distribution of the initial velocities $V_a$.

It is useful to consider some approximations. If the lens does not change neutron velocities too much and $\left(\delta/t\right)\left(\tilde{V}_a^2/\tilde{V}_b^2\right)\ll 1$, the correction to the calculated time of flight~\eqref{eq:eq21} takes the form $\xi(t)=-\left(b/\tilde{V}_b\right)\left(\delta/t\right)\left(\tilde{V}_a^2/\tilde{V}_b^2\right)$. Applying the relation $t=a/\tilde{V}_a$, we finally obtain
\begin{linenomath}\begin{equation}\label{eq:eq24}
\xi (t)=-\delta\left(b/a\right)\left(\tilde{V}_a/\tilde{V}_b\right)^3.
\end{equation}\end{linenomath}

If the time of flight is much longer than the period, as can be seen from Fig. 1, the neutron velocities captured by the lens do not differ too much from the average value $\langle\tilde{V}_a\rangle$, and the time dependence of the ratio $\tilde{V}_a/\tilde{V}_b$ can be neglected. In this approximation it is easy to estimate the duration of the pulse generated by the lens. Averaging~\eqref{eq:eq24} over the time interval equal to the duration of the neutron pulse $-\theta/2<\delta<\theta/2$, we obtain the pulse duration of the time image $\tau=-\theta\left(b/a\right)\left(\tilde{V}_a/\tilde{V}_b\right)^3$ and for time magnification it will be
\begin{linenomath}\begin{equation}\label{eq:eq25}
M=\frac{\tau}{\theta}=-\frac{b}{a}\left(\frac{\tilde{V}_a}{\tilde{V}_b}\right)^3.
\end{equation}\end{linenomath}

The negative sign takes into account the fact that the neutrons that left the source with a delay $\delta$ reach the observation point earlier than the calculated time. If the ratio $\tilde{V}_a/\tilde{V}_b$ is close to unity, we arrive at the estimate (15) given in the work \cite{Frank00}.

In fact, the class of time lenses is by no means limited to the case presented in Fig. 1, when the lens does not change the average neutron velocity and $\tilde{V}_a/\tilde{V}_b\approx 1$. In particular, the possibility of using a lens that decelerates neutrons was mentioned in \cite{Frank22}. The case of an accelerator lens is also quite possible. The scheme of operation of such devices is illustrated in Figure 3. It is obvious that when designing devices with such lenses, it is necessary to take into account the above-mentioned considerations of how they transform the shape and duration of the time pulse. At the same time, taking into account the factor $\left(\tilde{V}_a/\tilde{V}_b\right)^3$ included in~\eqref{eq:eq25} significantly limited the calculated value of the UCN density in the trap.

\begin{figure}[htb]
\centering
        \includegraphics[width=\linewidth]{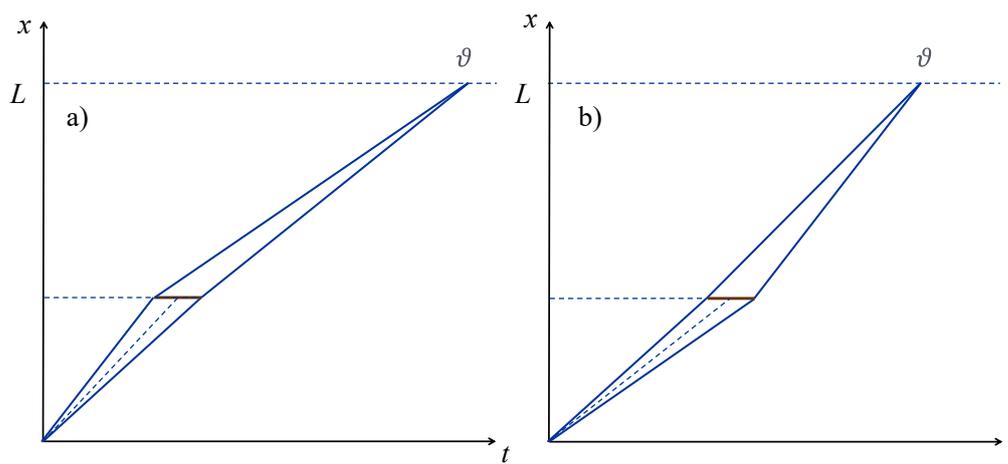}
\caption{ a) neutron decelerating lens and b) neutron accelerating lens.}
\label{fig:Fig3}
\end{figure}

It should be reminded that speaking about the duration of the pulse, we hereinafter mean the duration of the neutron bunch. When maintaining the average flux, this value directly determines the neutron density in the bunch, and in case of pulsed filling of the trap it also determines the neutron density in the latter.

\section{\uppercase{Time focusing methods.}}
\label{sec:sec4}
\subsection{\textit{Possible ways to transfer energy to a neutron}}
\label{sec:ssec4p1}

As noted above, the effect of a local time lens is based on the possibility of an exactly defined and time-variable energy transfer to a neutron. The two methods of energy transfer to a neutron based on the use of non-stationary quantum phenomena were considered in \cite{Frank00, Frank22}. This is the neutron spin flip by a resonant spin flipper, an obligatory component of which is the presence of an alternating magnetic field with a circular frequency $\Omega$, and the modulation of the fluх by some device providing amplitude or phase modulation of the flux with a frequency $\Omega$. In both cases, an energy equal to or being a multiple of the energy quantum $\Delta E=\hbar\Omega$ is transferred to the neutron. In \cite{Nesvizhevsky22} for this purpose it was proposed to use a method based on a change in the magnitude of the magnetic field over the time of neutron flight along this field region \cite{Niel89}.

When comparing different methods of transferring the desired amount of energy and momentum to a neutron, the following factors should obviously be taken into account: the maximum energy transferred by the lens, the efficiency of the lens, the ability to provide the necessary accuracy of the transferred momentum and the simplicity of quick changing of the amount of the energy transferred. The latter should change from its maximum to the minimum value during the period of the lens acting and return to the initial value again at the end of each cycle.

Let us consider the possible types of time lenses from this point of view.

\subsection{\textit{Lens based on non-stationary neutron diffraction}}
\label{sec:ssce4p2}

The physical basis of lenses of this type is a quantum phenomenon that occurs when a neutron flux is modulated by some device that provides an amplitude or phase modulation of the flux with a frequency $f=\Omega/2\pi$. The resulting state is a superposition of a large number of waves with different amplitudes $A_n<1$ and discrete frequency values of
\begin{linenomath}\begin{equation}\label{eq:eq31}
\omega_n=\omega_0+n\Omega,
\end{equation}\end{linenomath}
where $n$ is an integer \cite{Nosov91, FrankPAN94}. Out of the complete set of waves, only one has an energy corresponding to the necessary condition for time focusing at each given time moment. Therefore, the effectiveness of the lens is determined by the square of the modulus of the corresponding amplitude $|A_n|^2$. Assuming that the lens should change the neutron energy by an amount on the order of the UCN energy, $\Delta E=100$ neV, and implying that we use a first-order wave with $n=\pm 1$ as the resulting one, we obtain that the modulation frequency required for this should be $f=\Delta E /2\pi \hbar \approx 25$ MHz. It seems that such high-frequency modulation of a relatively wide beam is difficult to implement in practice. An alternative to this approach is a method in which the modulation of a neutron wave is achieved when some periodic structure moving across the direction of neutron propagation \cite{FrankPLA94}.

Let us assume that, generally speaking, the complex transmission function $H(x)$ of such a structure is periodic, $H(x)=H(x+d)$. Then, when such a structure moves along the $X$-axis at a velocity of $V_g$, a time modulation of intensity or phase with frequency 
\begin{linenomath}\begin{equation}\label{eq:eq32}
f=d/V_g,
\end{equation}\end{linenomath}
will take place at each point of the beam. The modulation frequency of several tens of MHz given above as an estimate is achieved, for example, if a structure with a period of several microns moves at a speed of several tens of meters per second. Thus, we are talking about a phase or amplitude diffraction grating moving across the beam.

In order to change the modulation frequency, a quasi-periodic grating can be used, the distance between the elements of which is a function of the coordinate. At a constant grating velocity, this leads to a time dependence on the frequency $f(t)$ and, accordingly, the transmitted energy $\Delta E(t)$. The characteristic time during which such a frequency change occurs should be large enough to provide a condition for quasi-periodic modulation of the flux. The formation of a discrete spectrum as in \eqref{eq:eq31} during the diffraction of UCNs on a moving quasi-periodic grating was observed in the experiment described in \cite{FrankPLA03}. The effect of time focusing at the UCN diffraction on a moving grating has also been experimentally demonstrated \cite{FrankJETPL03, Balashov04}. This approach to creating a time lens has been considered as the main one when discussing the possibility of creating a UCN source on a pulsed reactor \cite{Frank22}. This choice was influenced by the fact that the phenomenon of non-stationary UCN diffraction on a moving grating has been well studied both theoretically and experimentally  \cite{Bushuev16, KulinPRA16}.

The advantage of the method is its relative technical simplicity and relatively high diffraction efficiency for neutrons with a velocity of less than or of the order of 10-15 m/s. In addition, for a lens of this type, there seems to be no problem to quickly restore its initial state at the end of the focusing cycle.

However, the method has a significant drawback. The fact is that although the neutron energy changes by the value $\Delta E_n=\hbar\Omega_n=2\pi n\left(d/V_g\right)$ when the grating moves, this energy change corresponds to a change in the velocity component  directed along the beam only if the neutrons move strictly perpendicular to the grating plane.

In fact, UCNs propagating in a neutron guide have a velocity component $v_x$ perpendicular to the direction of their transportation. Therefore, the frequency of the flux modulation is $f=d/\left(V_g-v_x\right)$ and the energy transferred to the neutron depends on $v_x$, which is not a fixed value. Thus, in the presence of a transverse velocity $v_x$, the velocity $v_z$ after the lens differs from the estimated value required by the lens formulas~\eqref{eq:eq2} and~\eqref{eq:eq2a}, which results in an error.
\begin{linenomath}\begin{equation}\label{eq:eq33}
\Delta V_b\approx V_b\frac{v_x}{2V_g}.\;\;\;\left(v_x\ll V_g\right)
\end{equation}\end{linenomath}
According to this, there is an error in the time of flight.

In order for the pulse duration $\tau$ at the entrance to the trap to be significantly shorter than the flight time of the second section of the neutron guide $t_b$, it is necessary to ensure that the condition is met
\begin{linenomath}\begin{equation}\label{eq:eq34}
V_g\gg v_x\frac{t_b}{2\tau}.
\end{equation}\end{linenomath}

Since the transverse velocities of neutrons transported by neutron guides are usually several meters per second, condition~\eqref{eq:eq34} is hardly feasible if it is required that, with a neutron guide length of the order of several meters, the duration of the resulting time pulse is of the order of several milliseconds. But it was precisely these values that were discussed in \cite{Frank22} when discussing a UCN source with pulsed accumulation of neutrons.

\subsection{\textit{Magnetic lens with a resonant spin flip}}
\label{sec:ssec4p3}

Along with a time lens based on non-stationary UCN diffraction, \cite{Frank00} also considered the possibility of time focusing of neutrons based on a resonant spin flip of the neutron in a magnetic field. Resonant conditions are fulfilled when, in addition to the constant magnetic field $B$, an alternating field with Larmor frequency of $\omega_L=2\mu B/\hbar$ directed perpendicular to $B$ also affects the neutron. In this case the neutron energy changes by an amount $\Delta E=\hbar \omega_L$. To provide a given law of energy change $\Delta E(t)$ in \cite{Frank00}, it was proposed to use a slowly time-variable field $B(t)$ with a synchronous frequency change $\omega_L(t)$, which provides constant fulfillment of resonance conditions. The possibility of magnetic time focusing has found its experimental confirmation in the works \cite{Arimoto12, Imajo21}. However, the original idea was somewhat modified and instead of a time-variable field $B(t)$, a constant field  $B(x)$ depending on the coordinate was used, and the dependence of the field on time arose in the coordinate system associated with a moving neutron. The spin flip device was a gradient or adiabatic spin flipper \cite{Egorov74, Grigoriev97, Luschikov84} with the difference that the frequency of the high-frequency field varied in time, and the resonant conditions were fulfilled at slightly different points in space.

A magnetic lens based on this principle has an efficiency of 0.5, since it properly changes the energy only for neutrons with a single spin projection. To change the neutron velocity by a relatively small amount $\Delta V$, it is obviously necessary to change the neutron energy by
\begin{linenomath}\begin{equation}\label{eq:eq35}
\Delta E\approx m V \Delta V.
\end{equation}\end{linenomath}

If we are talking about UCNs, a several-T field is required for a noticeable change in their velocities. This is probably feasible, although today's experience in creating flippers with a strong field is limited to a value of about 1 T \cite{Arimoto12, Imajo21, Holley12}, which corresponds to an RF field frequency $f=\omega_L/2\pi$ of about 30 MHz.

An important issue is the velocity aperture of the lens set by the range of velocities with which it can operate. If the change in the neutron velocity $\Delta V$  is not too small compared to the initial one, the field $B$, in which the spin flip occurs, should also change by a noticeable amount. This means that the corresponding frequency of the high-frequency field should be changed within significant limits during the order of the period of the pulsed neutron source, as well as quickly restored to its initial state before the next bunch of neutrons arrives. It should be noted that the spin flip in such a device occurs at significantly different values of the magnetic field $B(x)$, that is, at different points relative to a long magnetic system with a magnetic field gradient. Thus, the ratio of the flight lengths before and after the lens also depends on the time and transferred energy.

An important feature of a magnetic lens based on a spin flip is that it can either only accelerate or only decelerate all neutrons. Therefore, when designing a UCN source, it is necessary to take into account the above-mentioned considerations about the shape and duration of the neutron pulse.

Another problem with this focusing method is that the energy transferred to the neutron is precisely determined by the frequency of the alternating field, rather than by the velocity component along the beam axis. Although the total energy of the neutron varies in a small region of space where the spin flip occurs, the change in the velocity of the neutron takes place throughout the entire path of the neutron movement in an inhomogeneous magnetic field. The force affecting the neutron is directed along the gradient of the magnetic field. However, in any extended area of space, the field gradient cannot have only one component directed along the beam and part of the energy is spent on changing the velocity component perpendicular to its axis. This leads to an error in the magnitude of the transferred longitudinal velocity component and, consequently, impairs the quality of time focusing.

\subsection{\textit{Magnetic lens with a time-variable magnetic field}}
\label{sec:ssec4p4}

Another possibility of changing the neutron velocity at time focusing is based on the use of a time-variable magnetic field available in some local region of space. Inside this region, the field is assumed to be homogeneous and, therefore, the force does not affect the neutron in this region. The change in energy and velocity occurs only at the boundaries of the region: when a neutron enters and exits it. However, if the magnitude of the magnetic field changes by an amount $\Delta B$ during the passage of this region, after passing the device, the neutron energy changes by an amount $\Delta E=\mu\Delta B$. The possibility of such an effect on the energy and velocity of the neutron was demonstrated in \cite{Niel89}.

Comparing this approach with the resonant spin flip method discussed above, one may note that its advantage is the relative simplicity of the required device, in which there is a single time-dependent field $B(t)$. In addition, unlike a lens with a resonant spin flip, the change in energy and velocity occurs in a fixed region of space.

However, the method has significant limitations. The need for a quick change of the magnitude of the magnetic field is in obvious contradiction with the desire to increase the magnitude of the magnetic field, which determines the time aperture of the lens. Similar to a lens based on a spin flip, a lens with a time-variable field can either only accelerate or only decelerate neutrons. The remarks concerning the non-defined connection between the transferred energy and the momentum transferred in the direction of the beam axis also remain valid.

\section{\uppercase{Using a neutron decelerating device to provide a pulsed beam structure}}
\label{sec:sec5}

As noted above, in \cite{Frank22} the concept of a UCN source on a pulsed reactor based on the use of a decelerating time lens was considered. The fact is that an increase in the initial velocity of neutrons can increase the intensity of the source. The important thing here is that the extraction of neutrons from the moderator-converter with higher velocities than those of the UCN allows one to use a more efficient converter and provides better conditions for transporting neutrons. In \cite{Frank22}, it was assumed that in addition to restoring the pulse structure of the beam, the lens should have reduced the longitudinal velocity of the neutron from 7-15 m/s to a value of 3-5 m/s.

An even more radical approach was proposed in \cite{Nesvizhevsky22}, where it was proposed to focus very cold neutrons (VCN) with velocities of about 50 m/s, followed by their deceleration in an escaping trap, as proposed in \cite{Summhammer86}. In  \cite{FrankPEPANL23}, the possibility of decelerating neutrons by a spin flip using an adiabatic flipper with a strong magnetic field was discussed. The structures of the neutron beam resulting from the use of these two neutron deceleration methods differ significantly.

\subsection{\textit{Local decelerating device with a fixed energy change value}}
\label{sec:ssec5p1}

One of the possible methods of neutron deceleration is based on a non-stationary spin flip in a magnetic field \cite{Drabkin60, Alefeld81}. The practical implementation of the method may consist in the use of an adiabatic or gradient spin-flipper \cite{Egorov74, Luschikov84, Grigoriev97} with a strong magnetic field. Passing through such a device, the energy of all neutrons changes by an amount $E_D=2\mu B$ where $\mu$ is the magnetic moment of the neutron and $B$ is the magnitude of the permanent magnetic field.

The energy change at the neutron spin flip in such a flipper was demonstrated in \cite{Weinfurter88}. It seems possible to create a flipper with a field of the order of 15-20 T, which is feasible in modern superconducting systems. The change in neutron energy will be 1.8 -2.4 $\mu$eV, which ensures that neutrons with initial velocities of 15-20 m/s are slowed down to UCN energy. Below, for short, we will call such a device a decelerator in order to avoid the term "moderator" widely used in neutron physics. We will assume that the slowing down of neutrons by the decelerator occurs in a relatively short section of their transportation in close proximity to the trap and following \cite{FrankPEPANL23} we will consider the question of the time structure of the beam of those neutrons that, after their slowing down by the decelerator, are able to be stored in the trap.

Before getting into the trap, neutrons must travel a noticeable path in the neutron guide, which we assume to be a mirror. The transverse velocity of neutrons $v_\perp$, normal to the surface of the neutron guide walls, is limited by the boundary energy value of its walls $E_{gd}$, so that $v_\perp<\sqrt{2E_{gd}/m}$. Thus, having fixed the maximum value of the total and transverse velocity of neutrons capable of being stored in a trap, we thereby have obtained a limit for the longitudinal velocity of such neutrons directed along the $Z$ axis of the neutron guide.
\begin{linenomath}\begin{equation}\label{eq:eq26}
v_{z\;fin}<\sqrt{V_{trap}^2-v_\perp^2}.
\end{equation}\end{linenomath}
Here $V_{trap}$ is the maximum velocity of neutrons capable of being stored in a trap. 

The decelerator, which forms the neutron flux immediately before neutrons enter the trap, changes the neutron energy by a certain amount $E_D$. If it is designed correctly, this change in kinetic energy is mainly due to a change in the longitudinal velocity of neutrons.

Since we are interested only in the distribution of the longitudinal velocity of neutrons and the kinetic energy associated with it, we will further on skip the z index of the values of our interest. Then the energy of neutrons entering the trap and being able to be stored in it lies in the range from zero to $E_{fin}=mv_{fin}^2/2$. Before deceleration, the energy of these neutrons should be in the range $E_D<E<E_D+E_{fin}$. At a sufficiently large energy value $E_D$, the energy range of neutrons that can be trapped after deceleration may be much smaller than the energy $\delta E\approx E_{fin}\ll E_D$  itself. But this means that the spread of the flight times $\delta t$ from the pulsed source to the decelerator and, accordingly, to the trap, of the "useful" neutrons of interest to us is not only small, but may be even shorter than the time of flight itself $t=L/V$ where $L$ is the length of the transport neutron guide
\begin{linenomath}\begin{equation}\label{eq:eq27}
\frac{\delta t}{t}=\frac{\delta V}{V}\simeq\frac{\delta E}{2E}\simeq\frac{E_{fin}}{2E_D}\ll1.
\end{equation}\end{linenomath}

Under favorable conditions, $\delta t$ can also be significantly less than the pulse repetition period of the reactor $T$. In this case, the flux of "useful" neutrons, which after deceleration will be converted to UCNs, will have a pulsed structure, although during their transportation through the neutron guide, the pulse duration will inevitably increase due to velocity dispersion $\delta V$.

It is easy to show that with the Maxwellian velocity distribution in the source, neutrons entering the trap directly from the source and neutrons obtained by being converted from VCNs to UCNs carry the same flux, but have a significantly different temporal and spatial structure. In the first case, the spread of the flight times $\delta t$ is much larger than the pulse repetition period $T$. In this case, the pulse structure practically disappears and an essentially uniform flux enters the trap, which corresponds to the average value of the neutron density. In the second case, when the duration of the bunch is significantly less than the pulse repetition period $\delta t/T\ll 1$, the length of the bunches is less than the distance between them. Accordingly, the neutron density in the bunch exceeds the average by a value of $G=T/\delta t$

It is also important that the range of velocities of "useful" neutrons $dV$ entering the decelerator is also considerably less than the required range of neutron velocities at the entrance to the trap. It follows directly from formula~\eqref{eq:eq27} that
\begin{linenomath}\begin{equation}\label{eq:eq28}
dV \simeq v_{fin} \frac{v_{fin}}{2V}.
\end{equation}\end{linenomath}

The latter circumstance significantly reduces the requirements for the parameters of the time lens if it is used together with a local decelerator. Let us recall that formulas ~\eqref{eq:eq27} and~\eqref{eq:eq28}  are obtained under the assumption that the energy of all neutrons is changed by the decelerator by the same amount.

\subsection{\textit{Decelerating device with a fixed velocity change value}}
\label{sec:ssec5p2}

Apart from the case of neutron deceleration with a fixed energy transfer, it is possible that the change in velocity rather than energy is fixed. In the latter case, the energy change depends on the initial velocity of the neutrons. Such a case was considered in \cite{Nesvizhevsky22}, where to slow neutrons down it was proposed to capture neutrons in a decelerating trap or to use the reflection of neutrons from a runaway potential barrier. It is assumed that initially the barrier moves at the speed of a neutron bunch, and at some moment it begins to slow down, reducing its speed to almost zero. Obviously, such a runaway barrier is equivalent to a decelerating trap in which neutrons being under the influence of force are reflected only from one wall of the trap. With this deceleration method, the range of neutron velocities $\Delta V$ and the length of the neutron bunch captured by the device are preserved, the latter is equal to
\begin{linenomath}\begin{equation}\label{eq:eq29}
S = \frac{\left(\Delta V\right)^2}{2w},
\end{equation}\end{linenomath}
where $w$ is acceleration. Since the velocity range $\Delta V$ does not change during deceleration, it is simply equal to the range of longitudinal speeds of the UCNs captured by the trap $\Delta V=\delta v_{fin}$. The pulse duration of neutrons slowed down in this way is
\begin{linenomath}\begin{equation}\label{eq:eq30}
\tau \approx \frac{S}{\delta v_{fin}}=\frac{\delta v_{fin}}{2w},
\end{equation}\end{linenomath}
while before the deceleration it was equal to $\Delta t\approx S/V$. It can be seen that with this method of deceleration, the duration of the bunch increases in proportion to the ratio of the initial and final velocities, and the density of neutrons in the bunch decreases accordingly.

Assuming, as it was done in \cite{Nesvizhevsky22}, for the velocity range the value $\delta v_{fin}=$ 7 m/s equal to the boundary velocity of neutrons reflected from the boundary of the magnetic field of $B=$ 4.3 T, and for acceleration $w=$ 100 m/s$^2$, we obtain from eq.~\eqref{eq:eq30} that $\tau=$ 35 ms. This value is significantly greater than the pulse duration of the sources mentioned above \cite{Ananiev77, Aksenov09, Garoby18, Chen16, Lopatkin21}.

\section{\uppercase{Combined effect of the time lens and the decelerator with fixed energy change value}}
\label{sec:sec6}

Although the use of a decelerator that changes the energy of all neutrons by a known value can, under favorable conditions, lead to partial preservation of the pulse structure of the neutron flux, the duration of the neutron bunches formed in this way will probably significantly exceed the initial duration of the neutron pulse. In this case, to reduce the duration of the bunches, a time lens can additionally be placed in the area between the source and the decelerator, reducing the duration of the bunch at the entrance to the decelerator. 

The joint use of a time focusing lens and a decelerator imposes restrictions on the lens parameters, since the velocity spectrum formed by the lens $V_b$,  is limited by the need for trapping neutrons after their slowing down by a decelerator.
\begin{linenomath}\begin{equation}\label{eq:eq36}
V_{min}\leq V_b\leq V_{max},\;\;\;V_{min}=\sqrt{2E_D/m},\;\;\;V_{max}=\sqrt{2\left(E_D+E_{fin}\right)/m}.
\end{equation}\end{linenomath}

\begin{figure}[htb]
\centering
        \includegraphics[width=\linewidth]{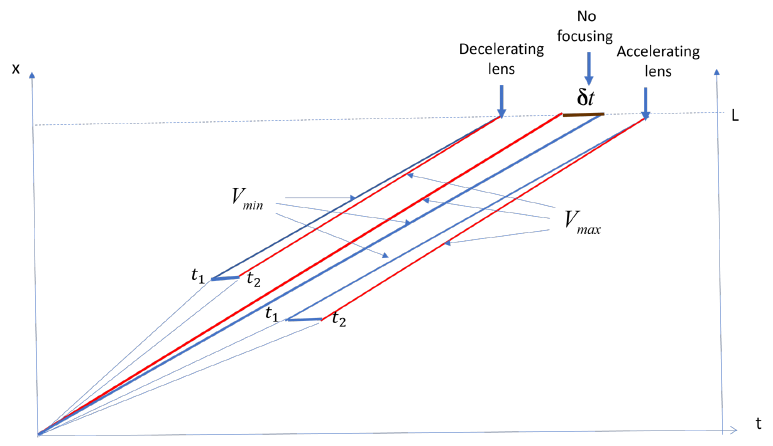}
\caption{Action of a time lens forming a neutron bunch at the entrance to the decelerator}
\label{fig:Fig4}
\end{figure}

Velocity diagrams explaining the peculiarities of focusing under these conditions are shown in Figure~\ref{fig:Fig4}. Limiting the variation range of velocities $V_b$ leaves the possibility of selecting only the values of the total flight time $\vartheta$ and the position of the lens $a$ as parameters. When choosing $\vartheta<t_{min}=L/V_{max}$, a lens should be decelerating, and in the case of $\vartheta>t_{max}=L/V_{min}$, it should be accelerating. The possibility of using a magneto-resonance  lens in the case $t_{min}<\vartheta<t_{max}$ is somewhat problematic, since the choice of such a flight time means that the slowing effect of the lens at some time moment must be changed to the accelerating one.

Since, the lens should is supposed to transform only the longitudinal component of the neutron velocity vz and the value of the transverse velocity component vx is limited by the properties of the neutron guide, the average neutron flux in all sections of the neutron guide is determined by the neutron density n and the range of longitudinal velocities
\begin{linenomath}\begin{equation}\label{eq:eq37}
f_z=\int^{v_2}_{v_1}{n\left(v_z\right) d v_z}.
\end{equation}\end{linenomath}
where $n\left(v_z\right)$ is the neutron density, determined by the total flux and the phase volume limited by the boundary velocity of the neutron guide. Assuming for simplicity that the neutron guide has a cylindrical cross section, we write down the expression for the neutron density in the form
\begin{linenomath}\begin{equation}\label{eq:eq38}
n\left(v_z\right)=\phi \int^{2\pi}_{0} {d \varphi} \int^{v_b}_{0} {v_r d v_r} \int^{v_z}_{0} {d V^{\prime}}.
\end{equation}\end{linenomath}
where $\phi$ is the normalizing factor proportional to the flux in the moderator and the boundary velocity of the neutron  guide. The initial velocity distribution is assumed to be Maxwellian and the neutron energy is small.

Because the lens cannot change the value of the average flux, the transformation of the velocity range by the lens leads to a change in the averaged over period neutron density in the beam. At the same time, the decelerating lens increases density, and the accelerating lens decreases it. Thus, the transformation coefficient of the neutron density is
\begin{linenomath}\begin{equation}\label{eq:eq39}
K=\frac{n_{fin}}{n}\frac{V_{max}-V_{min}}{a\left(t_1^{-1}-t_2^{-1}\right)}.
\end{equation}\end{linenomath}
The values of the time moments $t_1$ and $t_2$ are determined by a given range of final velocities~\eqref{eq:eq36}, the arrival time $\vartheta$ at the focus point and the flight length from the lens to the focus $b=L-a$
 \begin{linenomath}\begin{equation}\label{eq:eq40}
t_1=\vartheta-\frac{b}{V_{min}},\;\;\;t_2=\vartheta-\frac{b}{V_{max}}.
\end{equation}\end{linenomath}
 Substituting formulas~\eqref{eq:eq40} into~\eqref{eq:eq39} we get
\begin{linenomath}\begin{equation}\label{eq:eq41}
K=\frac{n_{fin}}{n}=\frac{\left(V_{max}^2-V_{min}^2\right)\left(\vartheta V_{min}-b\right)^2\left(\vartheta V_{max}-b\right)^2}{a^2\left[V_{min}^2\left(\vartheta V_{min}-b\right)^2-V_{max}^2\left(\vartheta V_{max}-b\right)^2\right]}.
\end{equation}\end{linenomath}

However, we are interested not in the period-averaged neutron density, but in the neutron density in the bunch, since here, as before, the filling mode of the trap is assumed to be pulsed. Therefore, the question of the bunch duration inversely proportional to the magnitude of the time magnification~\eqref{eq:eq25} becomes important. Omitting the negative sign in eq.~\eqref{eq:eq25}, which is insignificant in our case, we assume for approximate calculations
\begin{linenomath}\begin{equation}\label{eq:eq42}
M\approx\frac{b}{a}\left(\frac{v_2}{v_1}\right)^3,
\end{equation}\end{linenomath}
where $v_1=\left(V_1+V_2\right)/2$, $v_2=\left(V_{max}+V_{min}\right)/2$, $V_1=a/t_1$, $V_2=a/t_2$, and the times $t_1$ and $t_2$ are determined by formula~\eqref{eq:eq39}. Then the conversion factor of the pulsed neutron density that we are interested in is
\begin{linenomath}\begin{equation}\label{eq:eq43}
\Theta=K/M.
\end{equation}\end{linenomath}
An important issue determining the design of a lens is the amount of energy transferred by it. As can be seen from Figure~\ref{fig:Fig4}, for a decelerating lens, the maximum velocity change occurs for velocity $V_1$. Neutrons with such velocities are slowed down by the lens to a value determined by~\eqref{eq:eq36}. From this it follows that the maximum energy transfer by the decelerating lens is
\begin{linenomath}\begin{equation}\label{eq:eq44}
\delta E^{(1)}_{max}=E_D\left[\left(\frac{a}{\vartheta V_{min}-b}\right)^2-1\right].\;\;\;\left(a+b=L\right)
\end{equation}\end{linenomath}

Similarly, we obtain that the energy transfer by the accelerating lens is
\begin{linenomath}\begin{equation}\label{eq:eq45}
\delta E^{(2)}_{max}=\left(E_D+E_{fin}\right)\left[1-\left(\frac{a}{\vartheta V_{max}-b}\right)^2\right].\;\;\;\left(a+b=L\right).
\end{equation}\end{linenomath}

In principle, it is possible that the time $\vartheta$ of arrival at the focus point is located inside the interval $\delta t$, $L/V_{max}<\vartheta<L/V{min}$. In this case, the lens should accelerate some of the neutrons and decelerate other neutrons, which, apparently, excludes the possibility of using a single magnetic lens. However, it is quite possible to use two magnetic lenses, or a lens based on other principles. From the definition of the minimum and maximum velocities~\eqref{eq:eq36}, it follows that the maximum energy transfer by the lens in this case is equal $E_{fin}$, and in the case of using two lenses, it is even half as much.

The lens can also be useful in the case when, in addition to the dispersion of the neutron flight times to the decelerator $\delta t$, there is another source of increasing the duration of the bunch. Since the process of deceleration of neutrons is associated with their passage through an extended region with a magnetic field of the accelerator, and the deceleration time somehow depends on the speed, then such a source is seems to always exit. This creates time dispersion $\Delta t_{ds}$ that is added to the time-of-flight dispersion $\delta t$. 

\begin{figure}[htb]
\centering
        \includegraphics[width=0.95\linewidth]{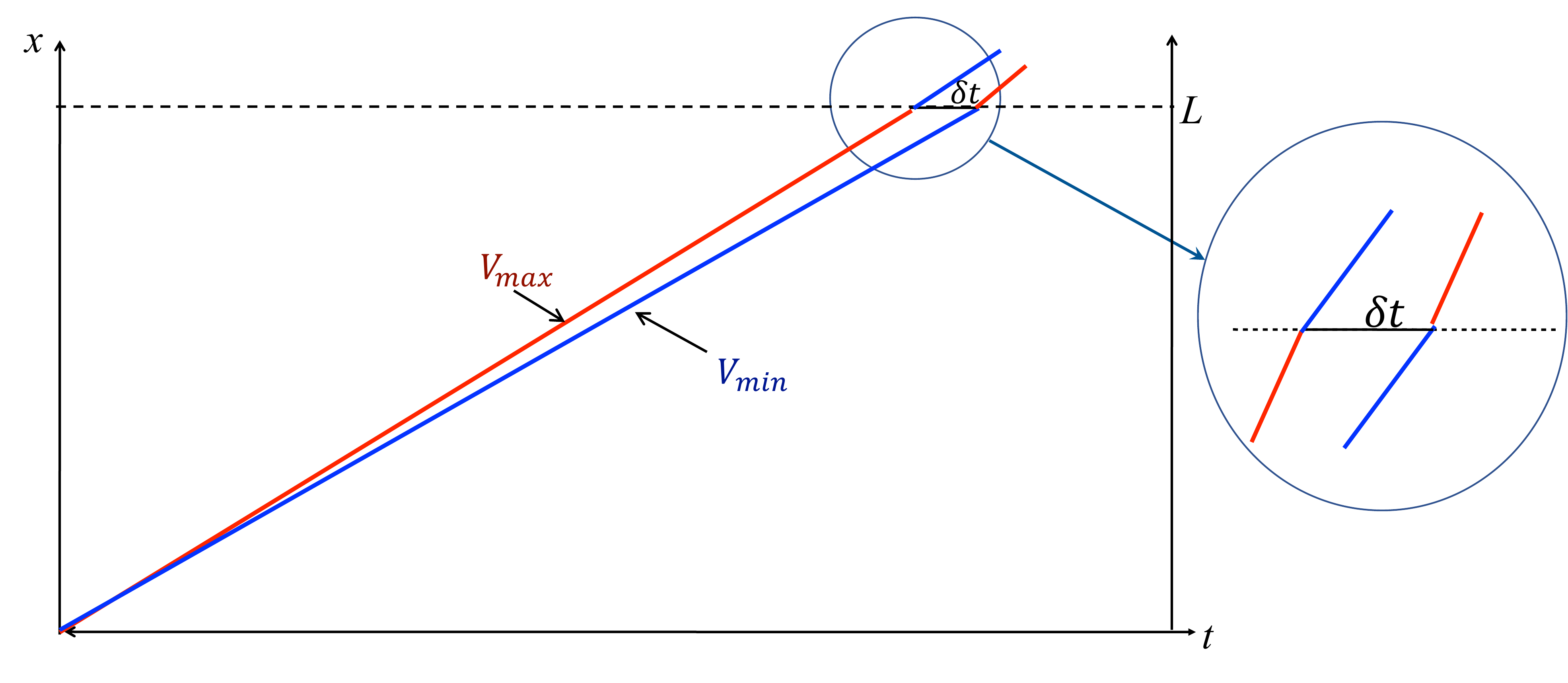}
\caption{Action of the inverting time lens}
\label{fig:FigInvertor}
\end{figure}

In this case, it may be useful to place the lens, or a combination of two lenses, directly in front of the decelerator. The final velocities should still be in the range from $V_{min}$ to $V_{max}$. But the neutrons that reach the decelerator at the moment $t_{min}=L/V_{max}$ (see Figure~\ref{fig:FigInvertor}) will now acquire the lowest possible velocity $V_{min}$ and spend the longest possible time on subsequent slowdown in the decelerator. On the contrary, the ones that get into the accelerator at the moment $t_{max}=L/V_{min}$ will acquire the maximum possible velocity $V_{max}$ and spend the least possible time on deceleration. As a result, the dispersion of the flight times $\delta t$ and the dispersion of the deceleration times $\Delta t_{ds}$ will partially or completely compensate for each other and the duration of the bunch will be determined by the difference between these time intervals.

\section{\uppercase{Problem of the pulsed valve}}
\label{sec:sec7}

One of the important problems in creating a UCN source with pulsed neutron accumulation is the problem of creating a fast valve at the entrance to the trap. Such a device should have a sufficiently high operation speed, and actually prevent leakage of neutrons from the trap.  The valve response time should be much shorter than the duration of the bunch, and the minimum possible neutron leakage can be easily estimated from the ratio~\eqref{eq:eq11}. It shows that in order to minimize the effect of the leakage through the valve on the time of neutron storage in the trap, the probability of neutrons passing through the valve $P$ in the closed state should be much less than the value $\mu\Sigma/s$.

Currently, there is a lot of experience in using shutters or valves for UCNs, the design of which is determined by their purpose. The first mentioning of the feasibility and usefulness of such devices can be found in \cite{Antonov69} and in the pioneering works by F.L. Shapiro’s team \cite{Groshev71}, where the possibility of storing UCNs in material traps was demonstrated. The purpose of UCN valves in experiments on storing UCNs in traps is to provide the possibility of filling the trap with neutron gas and to reliably isolate the trap for the duration of neutron storage and subsequent neutron release to the detector. The main requirement for such devices is the possible minimization of UCN leaks from the trap during storage. Since the storage time of UCNs in traps ranges from tens to many hundreds of seconds, the requirements for the operation speed of valves in such devices are quite mild.

For a long time, various types of mechanical valves, differing in their design, have been widely used both in experiments on measuring the lifetime of neutrons \cite{Kosvintsev86, Mampe89} and in the search for an electric dipole moment \cite{AltarevNPA80, Pendlebury84,Abel20}. Such valves provided sufficiently good insulation of the storage vessel, but the requirements for their speed were much milder than it is required in devices with pulsed density accumulation.

In UCN sources based on the spallation process \cite{Anghel09, Saunders13} mechanical valves are also used. Their purpose is to prevent the entering of UCNs into the moderator-converter, in which they are born, after the end of their generation cycle. Since such valves are part of a complex and expensive UCN source, the main requirement for their design is reliability and durability. The response time of such valves is determined only by the requirement of smallness compared to the duration of the neutron generation cycle, which is usually several seconds.
Devices for interrupting the UCN flux, designed for UCN time-of-flight spectrometry \cite{Fierlinger06, Bison23}, are much faster, because it is necessary to modulate the neutron flux according to a given time law. For a given flight base, the resolution of the spectrometer is determined by the effective duration of a single pulse in the classical time-of-flight method \cite{Fierlinger06, Bison23} or the frequency of beam interruption in correlation methods \cite{Novopoltsev10, Novopoltsev88}, including Fourier spectrometers \cite{KulinNIMA16}. Mechanical choppers provide quite a satisfactory operation speed, on the order of several milliseconds, but are not designed for sealed separation of areas with a different density of neutron gas.

Perhaps the only known type of valve that can be considered as a prototype for pulsed trap filling is the thin-film magnetic shutter proposed many years ago \cite{Novopoltsev88, Pokotilovski08}. This type of shutter consists of two or more alternating layers of ferromagnetic with a significantly different coercive force. Magnetically soft layers can be magnetized using an external magnetic field, while the magnetization of magnetically hard films should remain unchanged. At the same direction of magnetization, the system transmits neutrons, being at the same time a polarizer, while in the case of oppositely directed magnetization, the shutter reflects neutrons. The parameters of such shutters obtained in the works cited above are still far enough from those necessary for density accumulation traps.

Apparently, in the mentioned above case of using a flipper-decelerator located near the trap, there is one more possibility to create a pulse valve. To do this, the spin of the neutrons leaving the decelerator must be inverted, which can be done using an additional gradient or a resonant spin flipper. The latter acts only for a short time and turns off after a neutron bunch passes through it. Thus, polarized neutrons are stored in the trap, for which the strong magnetic field of the decelerator is impenetrable barrier \cite{Serebrov00}. Of course, during the operation of the flipper-valve, some of the neutrons inevitably leave the trap, being drawn into the magnetic field. But this situation is typical for the pulsed trap filling mode, which was taken into account when deriving formulas \eqref{eq:eq10}, \eqref{eq:eq12}. 
As noted above, the flipper reverses the spin of neutrons only in the presence of constant and orthogonal high-frequency (HF) magnetic fields. In this case, the HF field can be switched on according to a given time law.

\section{\uppercase{UCN source with resonant neutron deceleration and pulsed filling of the trap}}
\label{sec:sec8}
\subsection{\textit{Possible conception of the source}}
\label{sec:ssec8p1}

In the previous sections, some essential considerations that can be used as the basis for the UCN source with a pulsed filling of the trap were presented. Such a source does not yet exist and, when choosing its conception, there is an opportunity to choose from several variants. Let us briefly list them:
\begin{enumerate}
\item Extraction of UCNs from the converter, followed by their transportation and reconstruction of the pulse structure of the beam using a time lens. This approach, which goes back to the work \cite{Frank00}, was developed in \cite{Arimoto12, Imajo21}. A similar concept was suggested in \cite{Frank22}, where was proposed to accompany time focusing by a lens by neutron deceleration. For time focusing (rebunching), it was proposed to use a magnetic lens, or a lens based on non-stationary diffraction. The advantage of the method is its relative simplicity. The disadvantages are the loss of factor 2 on polarization and the low velocity aperture of the magnetic lens, in the first case, and limited diffraction efficiency and fundamental difficulties in creating a diffraction lens with the necessary parameters in the second.
\item Extraction of VCNs from the converter with subsequent time focusing and deceleration up to the UCN energy by capturing them into a decelerating trap \cite{Nesvizhevsky22}. The advantage of the method is the high efficiency of extraction and transportation of VCNs. The disadvantages are the difficulty of focusing the VCNs in the required velocity range, and the technical difficulty to capture and to transport neutrons in a decelerating magnetic or material trap.
\item Extraction of VCNs from the converter with their subsequent deceleration by the non-stationary neutron spin flip in a strong magnetic field using a flipper decelerator \cite{FrankPEPANL23}. The advantage of the method is a sufficiently high efficiency of extraction and transportation of VCNs. The pulsed structure is provided without using a rebunching device. The disadvantages are the loss of the factor of 2 due to polarization, the relatively long duration of the bunch when using a flipper with a magnetic field of 15-20 T, which is practically achievable at the present time, as well as the problem of the dispersion of deceleration times.
\item Extraction of VCNs from the converter, followed by their deceleration using a flipper decelerator complemented by magnetic compensator lenses. The advantage of the method is a significant reduction in the bunch duration compared to the option with a single decelerator and a very moderate value of energy transferred by the lenses, which makes their design easier. The disadvantages are the loss of the factor of 2 due to polarization and the complexity of the source design.
\item Extraction of VCNs from the converter, followed by their deceleration using a flipper-decelerator complemented with a focusing lens of a certain design. It differs from option 4) by an increase in the captured phase volume and, consequently, the intensity. An additional difficulty relates to a relatively large value of the energy transferred to neutrons by the focusing lens.
\end{enumerate}

\subsection{\textit{UCN source for the IBR-2M reactor}}
\label{sec:ssec8p2}

\begin{figure}[htb]
\centering
        \includegraphics[width=\linewidth]{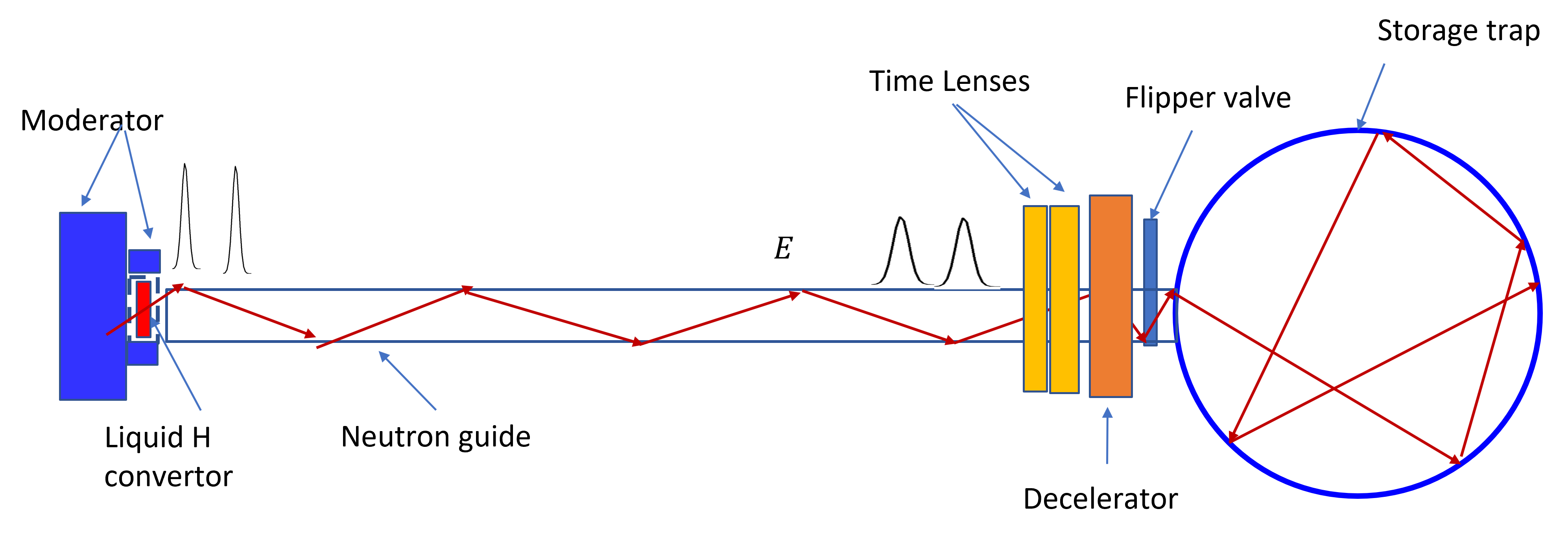}
\caption{Scheme of the designed UCN source at the IBR-2M reactor}
\label{fig:Fig5}
\end{figure}

When discussing the possible design of the source for the IBR-2M reactor, the choice was made in favor of the variant with a decelerator and compensating lenses. This option can provide a relatively short duration of the neutron bunch and at the same time seems technically feasible. The proximity of the magnetic system of the flipper-decelerator to the trap allows to use the flipper field as a barrier for neutrons stored in the trap. In combination with an additional pulse flipper, this allows us to solve the very difficult problem of a pulsed valve. Subsequently, such a source can be supplemented with another lens, allowing it to upgrade to option 5. The design of such source is shown in Figure~\ref{fig:Fig5}. Neutrons are generated in a thin moderator-converter located next to the main moderator of the reactor. After that neutrons are transported, using a mirror neutron guide, to two compensating lenses located just before the flipper-decelerator. There is an additional flipper in the decreasing field of the decelerator, which, together with the field of the decelerator, acts as a pulsed valve.

Technically complex elements of the source are a converter, compensating lenses, and a flipper decelerator with a valve flipper. The converter will serve as a flat vessel with liquid hydrogen, surrounded by a water moderator on three sides.

\begin{figure}[htb]
\centering
        \includegraphics[width=0.8\linewidth]{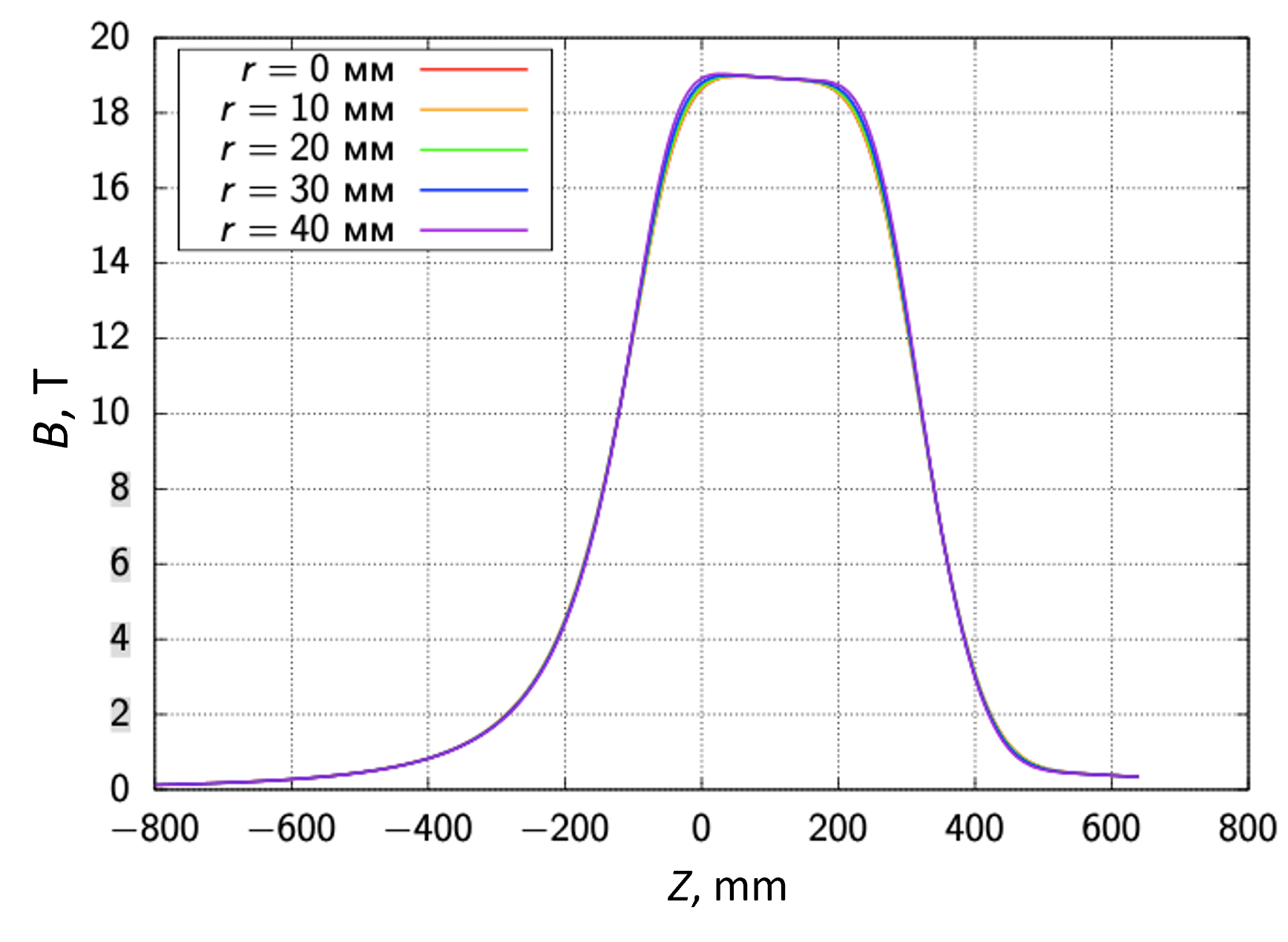}
\caption{ The dependence of the magnetic field on the coordinate on the axis of the magnetic system}
\label{fig:Fig6}
\end{figure}

A gradient spin flipper will be used as a decelerator. The magnetic field of the flipper will be created by a superconducting solenoid with a length of the current system of about 50 cm. The magnetic system is designed in such a way that in the region of the maximum magnetic field there will be a short, several centimeters long region with a magnetic field gradient of the order of 1 T/m. The magnitude of the magnetic field in the middle part of this region is about 18 T. The dependence of the magnetic field on the coordinate on the axis of the magnetic system is schematically shown in Figure~\ref{fig:Fig6}. The cryostat of the magnetic system will have a through hole with a diameter of 120 mm, in which a dielectric neutron guide with two high-frequency resonators of the "bird cage" type will be located. The resonator of the decelerator with an estimated frequency of about 520 MHz will be located in the above-mentioned region of a strong field gradient. The resonator will operate in the pulsed mode with a duty factor of about 20, so that the RF field is created only at the moment when a neutron bunch enters the decelerator. The second resonator, being part of the valve flipper, will be located in the output part of the magnetic system, where the magnetic field strength decreases to 0.4 T. It will also operate in the pulsed mode, creating an alternating magnetic field with a frequency of about 115 MHz.

The design of the storage trap has not yet been defined, but for preliminary estimates it was assumed that it has a shape close to spherical, and its inner surface is characterized by a boundary velocity of $V_g=6.9$ m/s and a loss parameter $\eta=10^{-4}$.  To reduce the dispersion in the time spent by neutrons on deceleration, it was decided to reduce the range of longitudinal velocities of UCNs entering the trap to a value of $3.5<v_z<5.5$ m/s.

For a given magnitude of the magnetic field $B=18$ T in the region of the spin flip, the change in the neutron energy in the decelerator is $\Delta E=2\mu B\approx2.2$ $\mu$eV. Hence the longitudinal (along the neutron guide) velocities of VCNs, which after deceleration correspond to the above-mentioned velocities of "useful" UCNs, lie in the range of $20.7<V<21.1$ m/s. At the expected length of the neutron guide $L=13$ m, the dispersion of the flight times of neutrons with such velocities is 13 ms.

Preliminary calculations show that the deceleration time dispersion is about the same, and the fastest neutrons, which spent the shortest time on transportation in the neutron guide, have the shortest deceleration time.  Therefore, in this case it may be beneficial to use compensating lenses, as discussed above in section \ref{sec:sec6}. Since such lenses decelerate the fastest neutrons to the minimum velocity and, conversely, accelerate the slowest ones to the maximum, the energy transferred by the lens to the neutron is equal to the difference in the kinetic energies of neutrons with maximum and minimum velocities. For the above-mentioned velocity range, it is about 95 neV. Neutrons loss or gain such energy by entering a magnetic field with a strength of 1.6 T. In the source, it is planned to use two compensating lenses of the Neil-Rauch type \cite{Niel89} with pulsed fields of this magnitude. The lenses will be placed one after the other in the area adjacent to the decelerator. One of them will be accelerating, and the other will be decelerating.

Apparently, it will not be possible to achieve ideal compensation for the duration of the bunch, this is why we assume that at the entrance to the trap their duration will be several milliseconds.
Under these assumptions, as well as with realistic parameters of the achievable surface roughness of the neutron guide, estimates of the possible source intensity were made. At the given limits of the longitudinal velocity of neutrons entering the trap and the full velocity of the UCNs stored in the trap, the dependence of the results on the value of the boundary velocity of the neutron guide turns out to be rather weak.

\begin{figure}[htb]
\centering
        \includegraphics[width=0.8\linewidth]{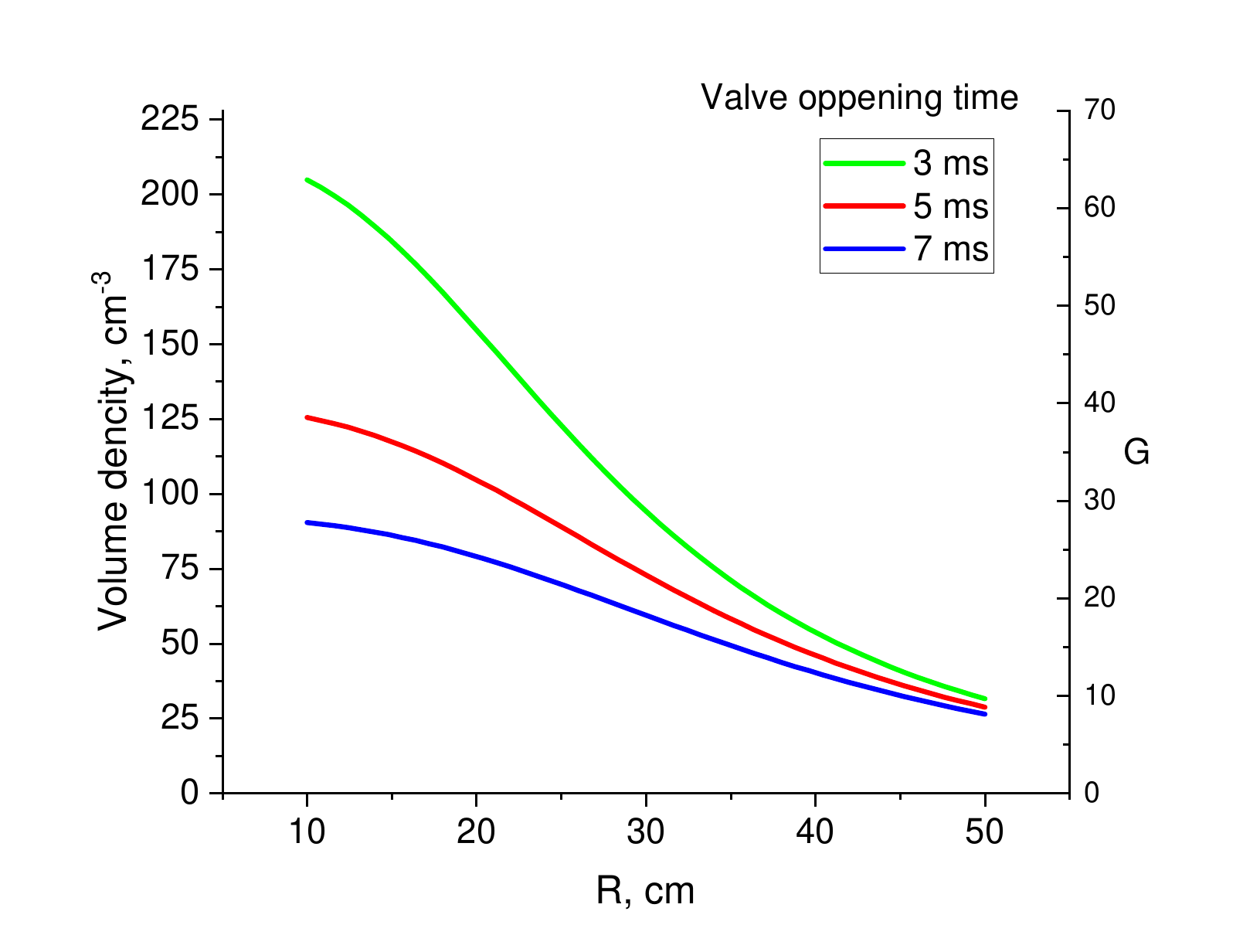}
\caption{Ratio of the flux in the trap to the average flux at its entrance; neutron flux in the trap and its dependence on the trap radius}
\label{fig:Fig7}
\end{figure}

The calculation results shown in Figure~\ref{fig:Fig7} indicate that with the chosen source conception on a pulsed reactor with a relatively low average power of 2MW, such as IBR-2M, the accumulation mode allows one to obtain a neutron density of about one hundred neutrons per cubic centimeter in a trap of a sufficiently large volume.

It should be noted that the choice of such a source conception was largely determined by the cost and possible project implementation period, and from the viewpoint of intensity it is not the best one. In particular, when complemented with an additional focusing lens, the density can be increased by 2-2.5 times, and the transition to a solid deuterium converter \cite{Granada09, Frei10, Doge21} will further increase the flux by about 30 times.

\section{\uppercase{Conclusion}}
The paper discusses various approaches to the creation of a UCN source based on the pulsed accumulation of neutron bunches in a trap. The latter makes it possible to accumulate UCNs in a trap with a density significantly higher than the density achievable by stationary filling of the trap with a time-averaged neutron flux. The main problem in creating such a source is the distance of the trap from the place of formation of initial neutron bunches with high neutron density. Several possible approaches to solve this problem have been considered. Along with the previously proposed principle of time focusing (rebunching) using a non-stationary time lens, considerable attention has been paid to an approach based on the use of a device called a decelerator, which changes the energy of all neutrons by a significant value directly near the trap. The possibility of a combined approach consisting in a combination of a decelerator and a time focusing has been discussed. 

In conclusion, the conception of a first stage UCN source, which is planned to be created at the IBR2-M pulsed reactor, has been described.

\section*{Acknowledgements}
Autors are grateful to E.V. Lychagin, A.Yu. Muzychka, V.N. Shvetsov and N.V. Rebrova for fruitful discussion.

The work was done according to Project 01-3-1146-3-2024/2028 of the JINR topical plan.


\label{sec:funding}
\section*{Funding}
Information on grants and other sources of financial support.

\label{sec:conflict}
\section*{Conflict of interest}
A conflict of interest is any relationship or area of interest that could directly or indirectly affect your work or make it biased, including financial relationships with organizations that sponsored the research or compensation received for consulting work.

\bibliographystyle{pepan}
\providecommand{\noopsort}[1]{}\providecommand{\singleletter}[1]{#1}%

\end{document}